\documentclass[preprint,showpacs,preprintnumbers,amsmath,amssymb]{revtex4}
\usepackage{graphicx}
\usepackage{dcolumn}
\usepackage{bm}
\usepackage{epsfig}
\usepackage{epstopdf}
\begin{document}
\title{Abelian-Higgs model from Cho-Faddeev-Niemi decomposition}
\author{A. Mohamadnejad}\altaffiliation {a.mohamadnejad@ut.ac.ir}
\author{S. Deldar}\altaffiliation {sdeldar@ut.ac.ir}
\affiliation{Department of Physics, University of Tehran, P.O. Box 14395/547, Tehran 1439955961,
Iran}
\begin{abstract}
Applying Cho-Faddeev-Niemi decomposition for SU(2) Yang-Mills field, we obtain the Abelian-Higgs Lagrangian by some approximation.
Abelian-Higgs Lagrangian with a spontaneous symmetry breaking potential has vortex solutions known as Nielsen-Olesen solutions.
We conclude that vortices as well as magnetic monopoles can exist in Cho-Faddeev-Niemi decomposition of SU(2) Yang-Mills field.

\end{abstract}


\maketitle

\section{\label{sec:level1}Introduction}

Decomposing the Yang-Mills field has been developed especially in the last decade. Although the original Yang-Mills theory is formulated in terms of the Yang-Mills
gauge field, this formulation is just suitable for studying the high energy dynamics of Yang-Mills theory where asymptotic freedom rules.
However, in the low energy limit we face the strong coupling problem, and its description in terms of the Yang-Mills field is no longer valid.
This assertion that the high energy and the low energy limits of a Yang-Mills theory characterize different phases is suggested especially by 't Hooft and Polyakov \cite{'t Hooft1,'t Hooft2}.
In investigating the low energy limit of the Yang-Mills theory, it is important to specify the most relevant degrees of freedom of the phenomenon.
Quark confinement as a typical phenomenon in the low energy regime caused by strong interactions is believed to be explained by topological defects including magnetic monopoles and vortices. Therefore, it is interesting to make another formulation of the Yang-Mills theory in terms of new variables reflecting the topological degrees of freedom.

Yang-Mills field decomposition was first proposed by Cho \cite{Cho1,Cho2} and has been developed by Faddeev and Niemi \cite{Faddeev,Niemi1,Niemi2} and Shabanov \cite{Shabanov1,Shabanov2}. This method enables one to explain and understand some of the low energy phenomena by separating the contributions of the topological defects in a gauge-invariant way.
In Cho method, the additional magnetic symmetry is introduced and it leads to a decomposition of the Yang-Mills field with six dynamical degrees of freedom.
One can show that magnetic monopoles appear in this decomposition and based on their condensations quark confinement is explained \cite{Deldar1,Deldar2} in the framework of the dual superconductor picture proposed in the 1970s \cite{Nambu1,Nambu2,Nambu3,Nambu4}.
Faddeev and Niemi study a special case of Cho decomposition which does not describe full QCD \cite{Evslin, Niemi3}. However, it leads to interesting results.
They also made an interesting conjecture that the Skyrme-Faddeev action can be interpreted as an effective action for the pure SU(2) Yang-Mills 
theory in the low energy limit \cite{Faddeev,Pak1,Pak2}. The Skyrme-Faddeev theory contains the topological knot solitons whose stability comes 
from the topological quantum number.

In this paper we show that Abelian-Higgs Lagrangian can be interpreted as an effective Lagrangian of  the Faddeev-Niemi's.
The Abelian-Higgs model is a field theoretical model with important applications in particle and condensed matter physics. It constitutes an 
appropriate field theoretical framework for the description of the phenomena related to superconductivity and its topological excitations are
 known as Abrikosov-Nielsen-Olesen vortices. At the same time, it provides the simplest setting for the mechanism of mass generation. The Nielsen-Olesen vortex is a topological excitation in the Abelian-Higgs model \cite{Nielsen}.
Vortices are solutions of the equations of motion in two dimensions, with a finite core size and a quantized magnetic flux. Therefore, we 
conclude that vortices can exist in the Faddeev-Niemi decomposition.

This paper is organized as follows: in sec.\ref{sec2} we review Cho-Faddeev-Niemi decomposition. In sec.\ref{sec3} we obtain 
Ginsburg-Landau Lagrangian from the Faddeev-Niemi Lagrangian by some approximations. Hence, Abelian-Higgs Lagrangian can be viewed 
as an effective Lagrangian of SU(2) Yang-Mills theory. Since Nielsen-Olesen vortex solution exists 
in Abelian-Higgs model, we conclude that vortices exist in the Cho-Faddeev-Niemi decomposition of SU(2) Yang-Mills field. Finally our conclusion comes in sec.\ref{sec4}.

\section{Decomposition of SU(2) Yang-Mills field} \label{sec2}

\subsection{Cho decomposition} \label{sec2.1}
In Cho decomposition of Yang-Mills field an isotriplet unit vector field $ \textbf{n} $ which selects the Abelian direction at each space-time is introduced. The Yang-Mills field
is restricted to the potential $ \widehat{\textbf{A}}_{\mu} $ which leaves $ \textbf{n} $ invariant:
\begin{equation}
\widehat{\textbf{A}}_{\mu} = A_{\mu} \textbf{n} + \frac{1}{g} \partial_{\mu} \textbf{n} \times \textbf{n}, \label{eq1}
\end{equation}
where
\begin{equation}
A_{\mu} = \widehat{\textbf{A}}_{\mu} . \textbf{n} \, \, \, , \textbf{n} . \textbf{n} = 1. \label{eq2}
\end{equation}
In eq. (\ref{eq1}) $ \widehat{\textbf{A}}_{\mu} $ is Cho projected potential which projects out the neutral gluons.
The above decomposition originally obtained by the following condition.
\begin{equation}
\widehat{D}_{\mu} \textbf{n} = \partial_{\mu} \textbf{n} + g \widehat{\textbf{A}}_{\mu} \times \textbf{n} = 0, \label{eq3}
\end{equation}
which means that the restricted Yang-Mills field is the field which leaves the field $ \textbf{n} $ invariant under the parallel transport.
In the low energy limit $ \widehat{\textbf{A}}_{\mu} $ dominates and it has a dual structure. In fact, the field strength tensor 
$ \widehat{\textbf{F}}_{\mu\nu} $  is decomposed into the electric field strength tensor $ F_{\mu\nu} $ and magnetic field strength tensor $ H_{\mu\nu} $:
\begin{equation}
\widehat{\textbf{F}}_{\mu\nu} = \partial_{\mu} \widehat{\textbf{A}}_{\nu} - \partial_{\nu} \widehat{\textbf{A}}_{\mu} + g \widehat{\textbf{A}}_{\mu} \times \widehat{\textbf{A}}_{\nu} = (F_{\mu\nu} + H_{\mu\nu})  \textbf{n}, \label{eq4}
\end{equation}
where
\begin{eqnarray}
F_{\mu\nu} = \partial_{\mu} A_{\nu} - \partial_{\nu} A_{\mu}, \label{eq5} \\ &&
\hspace{-43mm} H_{\mu\nu} = - \frac{1}{g} \textbf{n} . (\partial_{\mu} \textbf{n} \times \partial_{\nu} \textbf{n}). \label{eq6}
\end{eqnarray}
Note that singularities of $ \textbf{n} $ define $ \pi_{2} (S^{2}) $ which describe the non-Abelian monopoles. Indeed, one can obtain the Wu-Yang monopole \cite{Wu} by choosing the following ansatz
\begin{eqnarray}
A_{\mu} = 0 , \nonumber \\ &&
\hspace{-23mm} \textbf{n} = \widehat{\textbf{r}}= 
\begin{pmatrix}
\sin{\theta} \, cos{\varphi} \\
\sin{\theta} \, sin{\varphi} \\
\thickspace cos{\theta}
\end{pmatrix} . \label{eq7}
\end{eqnarray}

The restricted potential $ \widehat{\textbf{A}}_{\mu} $ contains two dynamical degrees of freedom for $ A_{\mu} $ corresponding to
 two polarizations and two topological degrees of freedom for $ \textbf{n} $. 

Although eq. (\ref{eq1}) which is relevant for describing the Abelian dominance \cite{Yotsuyanagi} and magnetic monopole dominance \cite{Stack1,Stack2}, plays an essential role in the infrared limit,
but one can extend it so that it contains all six on-shell degrees of freedom
\begin{equation}
\textbf{A}_{\mu} = \widehat{\textbf{A}}_{\mu} + \textbf{X}_{\mu} \label{eq9}
\end{equation}
where 
\begin{equation}
\textbf{X}_{\mu} . \textbf{n} = 0. \label{eq10}
\end{equation}
Eq. (\ref{eq9}) is Cho decomposition where the restricted potential $ \widehat{\textbf{A}}_{\mu} $ represents the neutral gluons and the valence potential $ \textbf{X}_{\mu} $ represents the colored gluons. The valence potential has 4 dynamical degrees of freedom. Therefore, $ \textbf{A}_{\mu} $ in eq. (\ref{eq9}) enjoys all six dynamical degrees of freedom of a SU(2) Yang-Mills filed.

Based on Cho decomposition (\ref{eq9}) the field strength tensor is
\begin{equation}
\textbf{F}_{\mu\nu} = \widehat{\textbf{F}}_{\mu\nu} + \widehat{D}_{\mu} \textbf{X}_{\nu} - \widehat{D}_{\nu} \textbf{X}_{\mu} + g\textbf{X}_{\mu} \times \textbf{X}_{\nu} . \label{eq01}
\end{equation}
and the Yang-Mills Lagrangian is expressed as
\begin{eqnarray}
L=-\frac{1}{4} \textbf{F}_{\mu\nu} .  \textbf{F}^{\mu\nu} \nonumber    \\ &&
\hspace{-28mm} =-\frac{1}{4} \widehat{\textbf{F}}_{\mu\nu} . \widehat{\textbf{F}}^{\mu\nu} 
 -\frac{1}{4} (\widehat{D}_{\mu} \textbf{X}_{\nu} - \widehat{D}_{\nu} \textbf{X}_{\mu}) . (\widehat{D}^{\mu} \textbf{X}^{\nu} - \widehat{D}^{\nu} \textbf{X}^{\mu}) \nonumber    \\ &&
\hspace{-24mm} - \frac{g}{2} \widehat{\textbf{F}}_{\mu\nu} . (\textbf{X}^{\mu} \times \textbf{X}^{\nu}) - \frac{g^{2}}{4} (\textbf{X}_{\mu} \times \textbf{X}_{\nu}) . (\textbf{X}^{\mu} \times \textbf{X}^{\nu}) \label{eq02}
\end{eqnarray}

The equation of motion that one obtains from the Lagrangian (\ref{eq02}) is the same as the original equation of the pure Yang-Mills theory:
\begin{equation}
\triangledown_{\nu} \textbf{F}^{\mu\nu} = 0 . \label{eq03}
\end{equation}
So Cho decomposition does not change the dynamics of QCD at the classical level. Note that Lagrangian  (\ref{eq02}) contains $ \textbf{n} $ explicitly, but the variation with respect to $ \textbf{n} $ does not create any new equation of motion. Therefore, $ \textbf{n} $ in Cho decomposition is not a dynamical variable.

Faddeev and Niemi proposed a special form of $ \textbf{X}_{\mu} $. In the following subsection Faddeev-Niemi decomposition which is a special form of Cho decomposition, is reviewed.

\subsection{Faddeev-Niemi decomposition} \label{sec2.2}
Faddeev and Niemi proposed a special form of $ \textbf{X}_{\mu} $ in Cho decomposition. In this proposal only two of four dynamical degrees of freedom is considered for $ \textbf{X}_{\mu} $. So it does not describe the full QCD. Consider the following identity
\begin{equation}
\textbf{A}_{\mu} = A_{\mu} \textbf{n} + \frac{1}{g} \partial_{\mu} \textbf{n} \times \textbf{n} + \frac{1}{g} \textbf{n} \times D_{\mu} \textbf{n} \label{eq11}
\end{equation}
if one restricts $ \textbf{A}_{\mu} $ so that $ D_{\mu} \textbf{n} = 0 $, the above identity reduces to Cho projection (\ref{eq1}).
However in the Faddeev-Niemi case, there is no such a restriction. Comparing Eq. (\ref{eq9}) and Eq. (\ref{eq11}), it is obvious that
\begin{equation}
\textbf{X}_{\mu} =   \frac{1}{g} \textbf{n} \times D_{\mu} \textbf{n} \, \, \, \, \, \Rightarrow  \textbf{X}_{\mu} . \textbf{n} =  (\frac{1}{g} \textbf{n} \times D_{\mu} \textbf{n}) . \textbf{n} =0. \label{eq12}
\end{equation}
Notice that one can construct an orthogonal three dimensional basis for the internal space by $ \textbf{n} $ and its derivatives $ \partial_{\mu} \textbf{n} $ and $ \textbf{n}\times\partial_{\mu} \textbf{n} $
\begin{equation}
 \textbf{n} . \partial_{\mu} \textbf{n} = \textbf{n} . (\textbf{n}\times\partial_{\mu} \textbf{n})
= \partial_{\mu} \textbf{n} . (\textbf{n}\times\partial_{\mu} \textbf{n}) = 0 . \label{eq13}
\end{equation}
Considering the above orthogonal basis and since $ \frac{1}{g} \textbf{n} \times \widehat{D}_{\mu} \textbf{n} $ is perpendicular to $ \textbf{n} $ one can expand it as
\begin{equation}
\textbf{X}_{\mu} =   \frac{1}{g} \textbf{n} \times D_{\mu} \textbf{n} = \frac{\phi_{1}}{g^{2}} \partial_{\mu} \textbf{n} + \frac{\phi_{2}}{g^{2}} \textbf{n}\times\partial_{\mu} \textbf{n}  \label{eq14}
\end{equation}
where the real scalar fields $ \phi_{1} $ and $ \phi_{1} $ are the coefficients of the expansion. Therefore, one gets
\begin{equation}
\textbf{A}_{\mu} = A_{\mu} \textbf{n} + \frac{1}{g} \partial_{\mu} \textbf{n} \times \textbf{n} + \frac{\phi_{1}}{g^{2}} \partial_{\mu} \textbf{n} + \frac{\phi_{2}}{g^{2}} \textbf{n}\times\partial_{\mu} \textbf{n}  \label{eq15}
\end{equation}
This is the Faddeev-Niemi decomposition. Note that two field degrees of freedom, $  \phi_{1} $ and $ \phi_{2} $, are added to the Cho projected potential. 
Now the variation with respect to $ \textbf{n} $ creates a new equation of motion \cite{Faddeev}. Therefore, Faddeev and Niemi interpret $ \textbf{n} $ as a a dynamical field. However, unlike Cho decomposition, 
the equations of motion that one obtains from the Faddeev-Niemi Lagrangian are not equivalent to original equations of pure Yang-Mills theory \cite{Evslin,Niemi3}.
Originally, Faddeev and Niemi achieve their decomposition by different argument and their main proposal was the completeness of their decomposition in four dimensions that has been criticized recently \cite{Evslin,Niemi3}.

Using Eq . (\ref{eq15}) in the definition of the SU(2) field strength tensor, $ \textbf{F}_{\mu\nu} = \partial_{\mu} \textbf{A}_{\nu} - \partial_{\nu} \textbf{A}_{\mu} + g \textbf{A}_{\mu} \times \textbf{A}_{\nu} $, one gets the following filed strength tensor for the Faddeev-Niemi decomposition
\begin{eqnarray}
\textbf{F}_{\mu\nu}=\lbrace F_{\mu\nu} +  (1-\frac{\phi_{1}^{2} + \phi_{2}^{2} }{g^{2}}) H_{\mu\nu} \rbrace \textbf{n}  \nonumber\\ &&
\hspace{-58mm} +\frac{1}{g^{2}} (D_{\mu} \phi_{1} \partial_{\nu} \textbf{n} - D_{\nu} \phi_{1} \partial_{\mu} \textbf{n} )  \nonumber\\ &&
\hspace{-58mm} +\frac{1}{g^{2}} (D_{\mu} \phi_{2} \textbf{n} \times \partial_{\nu} \textbf{n} - D_{\nu} \phi_{2} \textbf{n} \times \partial_{\mu} \textbf{n} ) , \label{16}
\end{eqnarray}
where the  definition of $ F_{\mu\nu} $ and $ H_{\mu\nu} $ is the same as Eq. (\ref{eq5}) and Eq. (\ref{eq6}), respectively. and the following 
equations are defined
\begin{eqnarray}
D_{\mu} \phi_{1} = \partial_{\mu} \phi_{1} - g C_{\mu} \phi_{2} \nonumber\\ &&
\hspace{-45mm} D_{\mu} \phi_{2} = \partial_{\mu} \phi_{2} + g C_{\mu} \phi_{1}. \label{eq17}
\end{eqnarray}
Then the Faddeev-Niemi Lagrangian is
\begin{eqnarray}
L=-\frac{1}{4} \textbf{F}_{\mu\nu} \,. \, \textbf{F}^{\mu\nu} \nonumber\\ &&
\hspace{-30mm} = -\frac{1}{4} [\lbrace F_{\mu\nu} +  (1-\frac{\phi_{1}^{2} + \phi_{2}^{2} }{g^{2}}) H_{\mu\nu} \rbrace \textbf{n}  \nonumber\\ &&
\hspace{-30mm} +\frac{1}{g^{2}} (D_{\mu} \phi_{1} \partial_{\nu} \textbf{n} - D_{\nu} \phi_{1} \partial_{\mu} \textbf{n} )  \nonumber\\ &&
\hspace{-30mm} +\frac{1}{g^{2}} (D_{\mu} \phi_{2} \textbf{n} \times \partial_{\nu} \textbf{n} - D_{\nu} \phi_{2} \textbf{n} \times \partial_{\mu} \textbf{n} )]^{2} \label{eq18}
\end{eqnarray}
Performing the variations with respect to new variables $ C_{\mu} $, $ \textbf{n} $, $ \phi_{1} $ and $ \phi_{2} $ one gets the following equations of motion
\begin{eqnarray}
\textbf{n} . \triangledown_{\nu} \textbf{F}^{\mu\nu} = 0 \nonumber\\ &&
\hspace{-28mm} \partial_{\mu} \textbf{n} . \triangledown_{\nu} \textbf{F}^{\mu\nu} = 0 \nonumber\\ &&
\hspace{-28mm} (\textbf{n} \times \partial_{\mu} \textbf{n}) . \triangledown_{\nu} \textbf{F}^{\mu\nu} = 0 \nonumber\\ &&
\hspace{-28mm} (D_{\mu} \phi_{1} - D_{\mu} \phi_{2} \textbf{n} \times) \triangledown_{\nu} \textbf{F}^{\mu\nu} = 0. \label{eq19}
\end{eqnarray}
One can show that the above equations allow the vortex solution \cite{Mo}. In the next section we provide one more clue to support the appearance of vortices.

Faddeev and Niemi obtain Skyrme-Faddeev Lagrangian from their decomposition.
This suggests that at low energies the physical states of the Yang-Mills theory are knotlike solitons.
In the next section we obtain Abelian-Higgs Lagrangian from the Fadeev-Niemi Lagrangian. This implies that vortices can appear in Faddeev-Niemi decomposition.

\section{Abelian-Higgs model as an effective Lagrangian of SU(2) Yang-Mills theory} \label{sec3}

In this section, to support the appearance of vortices,
we show that Abelian-Higgs Lagrangian can be interpreted as an effective Lagrangian of SU(2) Yang-Mills theory where the Yang-Mills field is decomposed via Eq. (\ref{eq15}).

Let us start with the Faddeev-Niemi Lagrangian (\ref{eq18}),
\begin{eqnarray}
L = -\frac{1}{4} \textbf{F}_{\mu\nu} \,. \, \textbf{F}^{\mu\nu} \nonumber\\ &&
\hspace{-28mm} = -\frac{1}{4} F_{\mu\nu} F^{\mu\nu} + \frac{1}{2g^{4}} ( \partial_{\mu} \textbf{n}.\partial_{\nu} \textbf{n} - \eta_{\mu\nu} \partial_{\lambda} \textbf{n}.\partial^{\lambda} \textbf{n} )
(D^{\mu}\phi)^{\ast}  (D^{\nu}\phi) \nonumber\\ &&
\hspace{-28mm} + \frac{i}{2g^{3}} H_{\mu\nu} (D^{\mu}\phi)^{\ast}  (D^{\nu}\phi)  - \frac{1}{2} H_{\mu\nu} F^{\mu\nu} (1-\frac{\phi ^{\ast} \phi}{g^{2}})  \nonumber\\ &&
\hspace{-28mm} - \frac{1}{4}  H_{\mu\nu} H^{\mu\nu} (1-\frac{\phi ^{\ast} \phi}{g^{2}})^{2} ,   \label{eq20}
\end{eqnarray}
where
\begin{eqnarray} 
\phi = \phi_{1} + i \phi_{2} ,  \nonumber\\ &&
\hspace{-28mm} D_{\mu}\phi = (\partial_{\mu}  + i g A_{\mu} ) \phi . \label{eq21}
\end{eqnarray}
The Lagrangian (\ref{eq20}) is invariant under the following local U(1) gauge transformations
\begin{eqnarray}
\phi \rightarrow e^{-i\alpha(x)}  \phi \nonumber\\ &&
\hspace{-30mm}  A_{\mu} \rightarrow A_{\mu} + \frac{1}{g} \partial_{\mu} \alpha(x) \label{eq22}
\end{eqnarray}
It is also invariant under rotations of $ \textbf{n} $ in three dimensional internal space that forms the non-Abelian group SO(3) which is a global symmetry.

One can eliminate the field $ \textbf{n} $ from the above Lagrangian by the following trick. Consider the vacuum fluctuations of the field $ \textbf{n} $ so that averaging over this field leads to the following equations
\begin{eqnarray}
\langle \partial_{\mu} \textbf{n}.\partial_{\nu} \textbf{n} - \eta_{\mu\nu} \partial_{\lambda} \textbf{n}.\partial^{\lambda} \textbf{n} \rangle = \eta_{\mu\nu} m^{2}   \nonumber\\ &&
\hspace{-40mm} \langle H_{\mu\nu} \rangle = 0  \nonumber\\ &&
\hspace{-44mm}   \langle H_{\mu\nu} H^{\mu\nu} \rangle = h^{2}  \label{eq23}
\end{eqnarray}
where $ m $ can be interpreted as the mass of the gauge boson $ A_{\mu} $ and $ h $ corresponds to the mass of the condensate field as we 
see shortly.
Note that the same technique is used in \cite{Faddeev} to get Skyrme-Faddeev Lagrangian. By the above approximation and 
changing the variable $ \phi $ 
\begin{eqnarray}
\phi \rightarrow \sqrt{2} \frac{g^{2}}{m} \phi \label{eq24}
\end{eqnarray}
the Lagrangian (\ref{eq20}) is
\begin{eqnarray}
L = -\frac{1}{4} F_{\mu\nu} F^{\mu\nu} + (D_{\mu}\phi)^{\ast}  (D^{\mu}\phi) - \frac{\lambda}{4} (\nu^{2} - \phi ^{\ast} \phi)^{2}   \label{eq25}
\end{eqnarray}
where
\begin{eqnarray}
\lambda = \frac{4h^{2} g^{4} }{m^{4}} \nonumber\\ &&
\hspace{-20mm} \nu^{2} = \frac{m^{2}}{2g^{2}} \label{eq26}
\end{eqnarray}
After spontaneous symmetry breaking and the following expansion of the field $ \phi $ around the vacuum $ \nu $
\begin{equation}
\phi = \nu + \frac{1}{\sqrt{2}}( \chi_{1} + i \chi_{2} ) \label{eq27}
\end{equation}
the Lagrangian (\ref{eq25}) is
\begin{eqnarray}
L = -\frac{1}{4} F_{\mu\nu} F^{\mu\nu} + \frac{1}{2} m_{A}^{2} A_{\mu}  A^{\mu} + \frac{1}{2}(\partial_{\mu}\chi_{1})^{2} + \frac{1}{2}(\partial_{\mu}\chi_{2})^{2}  \nonumber\\ &&
\hspace{-90mm}  - \frac{1}{2} m_{\chi}^{2} \chi_{1}^{2} + \sqrt{2} g \nu  A^{\mu} \partial_{\mu}\chi_{2} + cubic + quartic \, \, terms.  \label{eq28}
\end{eqnarray}
where
\begin{eqnarray}
m_{A} = \sqrt{2} \, g \nu = m \nonumber\\ &&
\hspace{-35mm} m_{\chi} = \sqrt{\lambda} \, \nu = \frac{g}{m}\sqrt{2} \, h  \label{eq29}
\end{eqnarray}
Therefore, $ m $ in  Eq. (\ref{eq23}) is the mass of the gauge boson $ A_{\mu} $ and the mass of the Higgs field $ \chi $ is proportional to the $ h $.

Finally, it is well known that the Abelian-Higgs model with Lagrangian  (\ref{eq25}) supports vortex solutions which are known as Nielsen-Olesen vortices \cite{Nielsen}.

\section{Conclusion} \label{sec4}
Decomposing the Yang-Mills field has been known for many years, but its physical significance has been realized recently.
This method is useful in the low energy limit where the perturbative methods fail. Quark confinement is an interesting phenomenon in this limit. It is believed that quark confinement
can be explained by topological defects such as magnetic monopoles and vortices. So it is natural to expect a decomposition for the Yang-Mills field that captures the topological degrees of freedom. For example, in Cho decomposition magnetic monopoles can appear and one can describe the confinement by their condensations. In this paper, we study a special form of Cho decomposition known as the Faddeev-Niemi decomposition and we show that in addition to magnetic monopoles, vortices can appear in this decomposition. For this purpose,
we show that Abelian-Higgs theory can be derived from SU(2) Yang-Mills theory where the Yang-Mills field is decomposed via eq. (\ref{eq15}). Using some approximations, we reach to the  Abelian-Higgs Lagrangian from the Faddeev-Niemi decomposition for the SU(2) Yang-Mills field. Therefore, vortices which are topological excitations
in the Abelian-Higgs model, can appear in Faddeev-Niemi decomposition.

\section{\boldmath Acknowledgments}
We are grateful to the research council of University of Tehran for supporting this study.

\end{document}